\documentclass[prb,amsmath,amssymb,twocolumn,floatfix]{revtex4}
\usepackage{graphicx}
\usepackage{dcolumn}
\usepackage{bm}
\usepackage{subfigure}
\usepackage{amsmath}
\usepackage{feynmf}
\usepackage{hyperref}

\newcommand{\ra}{\rightarrow}

\newcommand{\bs}{\boldsymbol}

\newcommand{\SRO}{Sr$_2$RuO$_4$}
\newcommand{\bk}{\boldsymbol k}
\newcommand{\bp}{\boldsymbol p}

\newcommand{\br}{\boldsymbol r}
\newcommand{\brp}{\boldsymbol r^\prime}

\usepackage{times}

\begin{document}
\title{Leggett modes and multi-band superconductivity in \SRO}
\author{Wen Huang$^1$, Thomas Scaffidi$^2$, Manfred Sigrist$^3$, and Catherine Kallin$^{1,4}$}
\affiliation{$^1$Department of Physics and Astronomy, McMaster University, Hamilton, Ontario, L8S 4M1, Canada}
\affiliation{$^2$Rudolf Peierls Centre for Theoretical Physics, Oxford OX1 3NP, United Kingdom}
\affiliation{$^3$Institut f\"ur Theoretische Physik, ETH-Z\"urich, CH-8093 Z\"urich, Switzerland}
\affiliation{$^4$Canadian Institute for Advanced Research, Toronto, Ontario M5G 1Z8, Canada}
\date{\today}

\begin{abstract}
\SRO~is a prototypical multi-band superconductor with three bands crossing the Fermi level. These bands exhibit distinct dimensional characteristics, with one quasi-2D $\gamma$-band and two quasi-1D $\alpha$- and $\beta$-bands. This leads to the expectation that the superconductivity on the $\gamma$-band may be only weakly Josephson-coupled to that on the other two bands. Based on an explicit microscopic weak coupling calculation appropriate for \SRO, we study the collective Leggett modes associated with the relative phase oscillations between the bands and show that a relatively soft Leggett mode exists due to the comparatively weaker inter-band Josephson coupling. These calculations also provide insight into why the superconducting gap magnitudes may be comparable on all three bands, despite the noticeable differences between the $\gamma$ and $\alpha/ \beta$ bands. The analyses can be readily applied to other multi-band superconductors. 

\end{abstract}

\maketitle
Multi-band superconductors possess physical properties that are not present in single-band superconductors. Depending on the nature of the interactions driving the Cooper pairing and the orbital character of the bands, the superconducting order parameter may not be dominated by one band with only much weaker induced superconductivity on the other bands. This is particularly so in multi-band systems with unconventional pairing symmetry, where the correlations underlying the superconductivity often involve electrons on different bands strongly interacting with each other. These inter-band interactions give rise to effective Josephson couplings between the superconducting order parameters of the different bands~\cite{leggett-mode}. As a consequence, in the ground state, the multiple order parameters are locked in a configuration with a particular set of relative phases and magnitudes. 

Under external perturbations or at finite temperatures, the relative phase between the multiple order parameters can fluctuate, costing a finite amount of energy that is determined by the inter-band couplings. These collective excitations are commonly referred to as Leggett modes.\cite{leggett-mode} They respond to electromagnetic fields in a peculiar manner, and are unlike the usual global U(1) phase fluctuations which are pushed up to the plasma frequency due to Coulomb interactions.\cite{Anderson:58} 

\begin{figure}
\includegraphics[width=5cm]{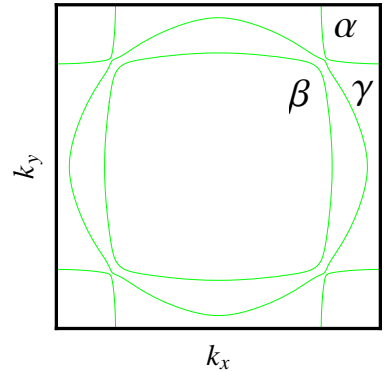}
\caption{The three Fermi surfaces of \SRO~in the $k_z=0$ plane. The $c$-axis dispersion is small and is ignored in our calculations.}
\label{fig:FS}
\end{figure}

The putative chiral $p$-wave superconductor \SRO~\cite{Mackenzie:03,Maeno:01,Kallin:09} is a prototypical multi-band system, with three bands crossing the Fermi energy -- two quasi one dimensional (1D) $\alpha/\beta$-bands and one quasi two dimensional (2D) $\gamma$-band (Fig.~\ref{fig:FS}).\cite{sro-review} The quasi-1D bands originate primarily from the hybridized 4$d$  $xz$ and $yz$-orbitals, while the $\gamma$-band is dominated by the $xy$-orbital. These orbitals are further mixed by spin-orbit coupling.\cite{Haverkort:08,Veenstra:14,Fatuzzo:15}  

The exact superconducting gap structure in this material is an ongoing debate.\cite{sro-review,chiral-review,Maeno:12} In spite of this, a few things can be said regarding the effective interactions between the low energy fermions on the three Fermi surfaces. Firstly, the intra-band Cooper pair scattering on the quasi-1D bands may be markedly different from that on the quasi-2D band. This could lead to one set of bands or the other dominating the superconductivity, as was first pointed out by Agterberg et al.\cite{Agterberg:97} However, as we will see, inter-band interactions make this less likely. Secondly, due to the quasi-1D nature of the $\alpha/\beta$-bands, the inter-band scattering between these two must be much stronger compared with that involving the $\gamma$-band. This naturally leads to a relatively weaker Josephson coupling between $\gamma$ and $\alpha/\beta$-bands. 

There is some experimental evidence in favor of comparable superconductivity on all three bands of \SRO,\cite{Nashizaki:00,davistunnel} and theoretically, a weak-coupling renormalization group analysis by Scaffidi {\it et al.}~\cite{Scaffidi:14} predicts comparable pairing strength on all of the bands in the parameter range believed to be appropriate to \SRO. However, this is an unresolved issue, and both experimental~\cite{Deguchi:04,Ray:14} and theoretical~\cite{Zhitomirsky:01,Nomura:02,Raghu:10,Garaud:12,Huo:13,Wang:13,Tsuchiizu:15} indications exist in support of a state where one of the two sets of bands dominates. 

Zhitomirsky and Rice~\cite{Zhitomirsky:01} have studied the effects driven by the inter-band interactions in a simplified two-band model, using phenomenological estimates for the interactions.  There, it was found that a reasonable amount of inter-band interaction is necessary to bind together the primary and the passive superconducting bands. In this work, we evaluate the effective inter-band interactions and Josephson couplings in \SRO~via explicit microscopic calculations following Scaffidi et al.\cite{Scaffidi:14} We will show that a relatively soft Leggett mode should be present because of the comparatively weaker coupling between the quasi-2D $\gamma$- and quasi-1D $\alpha/\beta$-bands. This detailed investigation into the inter-band interactions also helps to elucidate how the three bands may or may not support comparable Cooper pairing. 

In a chiral $p$-wave superconductor, whether single-band or multi-band, additional phase modes may also arise in connection with the relative phase fluctuations between the two chiral components. These are referred to in literature as ``clapping" modes.\cite{Wolfle:77} While we do not study these modes in detail, we argue that some of their experimental signatures differ from the Leggett modes and the two types of collective modes may be distinguished. 
 
Finally, although there is strong evidence for the time-reversal symmetry breaking (TRSB) chiral $p$-wave superconductivity in \SRO,\cite{Luke:98,Ishida:98,Duffy:00,Nelson:04,Xia:06} difficulties remain in reconciling the expectations for this order and a few key experiments, including the puzzling absence~\cite{Matsumoto:99,Kirtley:07,Hicks:10,Curran:14} (or smallness~\cite{Liu:16}) of spontaneous edge current (although recent years have seen numerous attempts to explain the absence of edge current\cite{Imai1213,Lederer:14,Bouhon:14,Scaffidi:15,Tada:15,Huang:14,Huang:15}), the anomalous suppression of the in-plane upper critical field $H_{c2}$\cite{Deguchi:02} and indications of a first order transition for in-plane $H_{c2}$  at low temperatures\cite{Yonezawa:13} (see Ramires et al.\cite{Ramires:16} for a recent attempt to explain this). It is thus tempting to ask whether \SRO~could in fact support an alternative TRSB odd-parity superconducting order, made possible by the multi-band nature,\cite{Agterberg:99} analogous to what has been discussed in the context of MgCNi$_3$,\cite{Voelker:02} some iron-based superconductors,\cite{Stanev:10,Lin:12,Maiti:13} and SrPtAs.\cite{Biswas:13} There, TRSB is associated with a complex order parameter configuration on three or more bands. In relation to \SRO, a helical $p$-wave pairing with complex multi-band order parameter for example would seem consistent with the experiments mentioned above. Note that a one-band helical $p$-wave is intrinsically time-reversal invariant, in the sense that the spin up and down electrons form Cooper pairs of opposite orbital angular momenta.\cite{Wolfle:77} However, in our model, we find that \SRO~lacks the ingredients favorable for the formation of this type of TRSB multi-band superconductivity.

The rest of the paper is organized as follows. We first formulate in Sec~\ref{sec:Model} a qualitative description of the multi-band superconductivity in \SRO, and then substantiate in Sec~\ref{sec:WeakCoupling} with inputs obtained from microscopic weak coupling calculations. We then make specific analyses of the Leggett modes, along with a discussion of the experimental consequence in connection with Raman scattering in Sec \ref{sec:Raman}. Finally we examine the possibility of exotic TRSB multi-band chiral and helical pairings in Sec \ref{sec:TRSB}. 

\section{Effective Model}
\label{sec:Model}
As is clear from the above discussion, a study of the multi-band dynamics in \SRO~requires a knowledge of the Josephson couplings between the multiple order parameters on the three bands. We start here by introducing an effective model to qualitatively capture the main features of the inter-band couplings.  Despite the lack of microscopic accuracy, this model is instructive for understanding the properties of the ground state and the collective phase modes. 

The effective Hamiltonian may be written as,
\begin{eqnarray}
&&H = \sum_{\mu,\sigma} \int d\br \psi^\dagger_{\mu,\sigma}(\br)(\frac{\hat{\bp}^2}{2m_\mu}-\mu)\psi_{\mu,\sigma}(\br) \nonumber \\
   && +\sum_{\substack{\mu,\nu\\ \sigma,\sigma^\prime}} \int d\br \psi^\dagger_{\mu,\sigma}(\br)\psi^\dagger_{\mu,\sigma^\prime}(\brp)V^{\mu\nu}_{\sigma\sigma^\prime}(\br-\brp) \psi_{\nu,\sigma^\prime}(\brp) \psi_{\nu,\sigma}(\br) \,. \nonumber \\
&& 
\label{eq:Hamiltonian}
\end{eqnarray}
Here $\mu = \alpha,\beta,\gamma$ are the band indices, $m_\mu$ is the effective mass for band-$\mu$, and $\sigma$ represent pseudospins which differ from the original spin indices due to spin-orbit coupling. The second term describes the {\it effective} electron-electron interactions between band-$\mu$ and band-$\nu$ in a particular {\it presumed} pairing channel (different channels are characterized by different effective interactions). These interactions presumably originate from Coulomb correlations and their associated particle-hole density-wave fluctuations. 

Note that we are only considering intra- and inter-band interactions which scatter pairs of electrons, respectively, within a band and from one band to another. Effectively, this amounts to having no inter-band Cooper pairs. No particular forms are specified for the interactions $V^{\mu\nu}(\br-\brp)$ at this point.  However, it is assumed that such interactions lead to the highly anisotropic chiral $p$-wave pairing with comparable pairing amplitudes on all of the bands, as found in earlier calculations.\cite{Scaffidi:14} In particular, the inter-band interactions $V^{\alpha\gamma}$ and $V^{\beta\gamma}$ are considered weak compared to $V^{\alpha\beta}$ as well as to the intra-band interactions, while $V^{\alpha\beta}$ is relatively strong and may even exceed $V^{\alpha\alpha/\beta\beta}$ due to the quasi-1D nesting.\cite{magnetic-susc} One may make a further simplifying approximation and set $V^{\alpha\alpha} \simeq V^{\beta\beta}$ on account of the similarity of the two 1D band structures.

As is elaborated in Appendix \ref{app1}, the eigenvectors of the interaction matrix $\hat{V}$ qualitatively approximate the possible order parameter configurations in the {\it presumed} pairing channel. In particular, the eigenstate with the most attractive eigenvalue corresponds to the most favorable configuration. 


After a Hubbard-Stratonovich transformation in the Cooper channel using auxiliary fields $\Delta_\mu$ (the superconducting order parameter) and integrating out the fermionic fields, the effective action becomes, 
\begin{equation}
S= \int d\tau d^2r\left( \sum_{\mu,\nu}\Delta_\mu^\ast \hat{\mathcal{V}}_{\mu\nu} \Delta_\nu - \sum_\mu \text{Tr~ln}G^{-1}_\mu \right)  \,,
\label{eq:action}
\end{equation}
where the first term may be simplified to $(\hat{\Delta}^\ast)^T \hat{\mathcal{V}} \hat{\Delta}$ with $\hat{\Delta} = (\Delta_\alpha,\Delta_\beta,\Delta_\gamma)^T$ and $\hat{\mathcal{V}} = -\hat{V}^{-1}$, and the Gor'kov Green's function is given by 
\begin{equation}
\hat{G}^{-1}_\nu= -\begin{pmatrix}
\partial_\tau -\frac{\nabla^2}{2m_\nu}-\mu_\nu &  -\Delta_\nu \\
-\Delta_\nu^\ast & \partial_\tau + \frac{\nabla^2}{2m_\nu}+\mu_\nu 
\end{pmatrix}  \,. 
\label{eq:Greenfunction}
\end{equation}
In this action we have ignored the vector potential which is irrelevant to our discussion. 

Particular attention is due for the coupling matrix $\hat{\mathcal{V}}$, whose off-diagonal elements describe inter-band Josephson couplings. Our expectation of much weaker $V^{\alpha\gamma}$ and $V^{\beta\gamma}$ compared to $V^{\alpha\beta}$ as well as the other interactions immediately leads to interband couplings of similar nature. On these bases, we take,
\begin{equation}
\hat{\mathcal{V}}= \frac{1}{V_0}\begin{pmatrix}
a_1 & \lambda & \eta_1 \\
\lambda & a_2 & \eta_2 \\
\eta_1 & \eta_2 & a_3 
\end{pmatrix}  \, .
\label{eq:Vinverse}
\end{equation}
with $|\eta_1|,~|\eta_2|\ll |\lambda|$. Here $V_0$ sets the overall interaction energy scale, the quantities $\lambda$ and $\eta_1 / \eta_2$ describe respectively the inter-band $\alpha$-$\beta$ and $\gamma$-$\alpha/\beta$ Josephson couplings, while the $a_i$'s are the intra-band couplings irrelevant to the rigidity of the relative phases between the bands. Noting the similarity between the quasi-1D bands, one may approximate $|\eta_1| \simeq |\eta_2| = \eta$.

The rigidity of the relative phase between the bands is determined by the inter-band couplings. Assuming the same order parameter amplitude on all bands, and setting $\lambda>0,~\eta_1\sim\eta_2 = \eta<0$ in light of our numerical results to be presented in the next section, our analyses follow the standard procedure~\cite{Lin:12,Ota:11} and are given in detail in Appendix \ref{app:Leggett} (see a more thorough derivation in Marciani et al.\cite{Marciani:13}). To simply quote the main conclusion: the system exhibits a relatively soft Leggett mode, with an excitation gap that is determined by the inter-band couplings $\eta$ in the following way, 
\begin{eqnarray}
w_{L}&=& \sqrt{\frac{3|\eta|}{N_0 V_0} }\Delta_0 \,,
\label{eq:LeggettwL}
\end{eqnarray}
where for simplicity we have assumed similar density of states $N_0$ and gap amplitudes on all bands. This mode is a consequence of phase fluctuations on the $\gamma$-band with respect to the other two bands. In the limit of vanishing interaction between the two sets of bands, this mode becomes massless. 

The particular set of inter-band couplings considered above is free of frustration (see Sec~\ref{sec:TRSB}). On the other hand, a set of frustrated inter-band interactions not realized in our model of \SRO, such as one that would lead to $\lambda>0,~\eta_1 > 0~\text{and}~\eta_2<0$, gives rise to an anomalously soft Leggett mode, as has been shown previously~\cite{Lin:12,Ota:11} (see Appendix \ref{app:Leggett}).

\section{weak coupling calculations}
\label{sec:WeakCoupling}
We now present a microscopic calculation of the interaction matrix $\hat{V}$ for \SRO. The first step is to obtain the effective band interactions using the microscopic Hamiltonian of the three Ru $t_{2g}$ 4$d$-orbitals. This can be achieved following the weak coupling renormalization group calculations by Scaffidi {\it et al.}\cite{Scaffidi:14} of the effective interaction $V^{\mu\nu}(\bs k,\bs p)$ associated with each Cooper pair scattering process on any pair of $\mu$- and $\nu$-band Fermi surfaces. For the sake of brevity, we refer to Ref.~\onlinecite{Scaffidi:14} for details and only sketch the calculations here. 

Most crucially, the study starts with on-site Coulomb interactions in the orbital basis,
\begin{eqnarray}
H_\text{int}&=& \sum_{i,a,s\neq s^\prime}\frac{U}{2} n_{ias} n_{ias^\prime} + \sum_{i,a\neq b,s,s^\prime} \frac{U^\prime}{2}n_{ias}n_{ibs^\prime} \nonumber \\
&+& \sum_{i,a\neq b,s,s^\prime} \frac{J}{2} c^\dagger_{ias}c^\dagger_{ibs^\prime}c_{ias^\prime}c_{ibs} \nonumber \\
&+& \sum_{i,a\neq b,s\neq s^\prime} \frac{J^\prime}{2}c^\dagger_{ias}c^\dagger_{ias^\prime}c_{ibs^\prime}c_{ibs} \,,
\end{eqnarray}
where $i$ is the site index, $a=xz,yz,xy$ is the orbital index, $s$ denotes the spin, $n_{ias} \equiv c^\dagger_{ias}c_{ias}$, $U^\prime=U-2J$, and $J^\prime=J$ where $J$ is the Hund's coupling. Following Raghu {\it et al.},\cite{Raghu:10} these interactions are treated perturbatively in the limit $U,J\ll W$ where $W$ is the bandwidth. Thus $J/U$ fully parameterizes the interactions in the model. Projecting all interactions to the Fermi level, $V^{\mu\nu}(\bk,\bp)$ in the Cooper channel is then evaluated up to the one-loop level, as is appropriate in the weak coupling limit. Finally, the superconducting gap function is obtained by solving the linearized gap equation using $V^{\mu\nu}(\bk,\bp)$. 

For a range of interaction and tight-binding parameters thought to be appropriate for \SRO, an anisotropic chiral $p$-wave pairing emerges as the most attractive solution to the gap equation (although a helical pairing represents a close competitor, see also Sec \ref{sec:TRSB}). We denote this gap in the following form,
\begin{equation}
\hat{\Delta}_{\bs k} = \begin{pmatrix}
 \Delta_{0\alpha} \phi_\alpha(\bs k) \\
 \Delta_{0\beta} \phi_\beta(\bs k) \\
 \Delta_{0\gamma} \phi_\gamma(\bs k) 
\end{pmatrix} \,,
\label{eq:GS}
\end{equation}
where $\phi_\mu(\bs k)$ is the normalized form factor of the full anisotropic chiral $p$-wave gap function on band-$\mu$, and the vector $\hat{\Delta}=( \Delta_{0\alpha}, \Delta_{0\beta}, \Delta_{0\gamma})^T$, with its elements indicating the relative phase and magnitude of the order parameters on the three bands, specifies the order parameter configuration. Note that these anisotropic pairing gaps in general lead to noticeably reduced edge current,\cite{Bouhon:14,Scaffidi:15,Huang:15} with strong further suppression when combined with surface disorder.~\cite{Lederer:14,Scaffidi:15}. Similarly anisotropic gaps on the two quasi-1D bands have also been invoked to explain tunneling conductance along the $c$-axis.\cite{davistunnel}

In Appendix \ref{app1} we formulate an approach to extract the effective intra- and inter-band interactions.  Essentially, in analogy to Scalapino {\it et al.}~\cite{Scalapino:86} formulated for a one-band model, the integrated inter-band interaction is approximated by,
\begin{equation}
V^{\mu\nu} = \frac{\oint_{\mu FS}d\bk \oint_{\nu FS}d\bp \frac{\phi^\ast_\mu (\bs k) V^{\mu\nu}(\bs k, \bs p) \phi_\nu (\bs p)}{v_\mu(\bs k) v_\nu(\bs p)}}{ \left( \oint_{\mu FS} d\bk \frac{|\phi_\mu (\bk)|^2}{v_\mu (\bk)}  \right)^{\frac{1}{2}} \left( \oint_{\nu FS} d\bp \frac{|\phi_\nu (\bp)|^2}{v_\nu (\bp)}  \right)^{\frac{1}{2}} } \,, 
\label{eq:V0maintext}
\end{equation}
where $v_{\mu}(\bs k)$ is the $\mu$-band Fermi velocity at Fermi wavevector $\bk$. For the parameters $J/U=0.06$, $\lambda_\text{SOC} = 0.1t$,\cite{SOCnote} used in Scaffidi {\it et al.}~\cite{Scaffidi:14} ($t$ is the primary in-plane hoping intergral of the 1D orbitals),  we obtain,
\begin{equation}
\hat{V} = V_0
\begin{pmatrix}
    0.5206&   -1.2181 &  -0.0635\\
   -1.2181&    0.3427 &  -0.0608\\
   -0.0635&   -0.0608 &  -1.0000
\end{pmatrix}
\label{eq:V0matrix}
\end{equation}
where $V_0>0$ sets the overall interaction energy scale in the pairing channel under consideration. Most notable features of this matrix include: a rather strong interaction between the two quasi-1D bands, considerably weaker inter-band interactions between the quasi 1D and 2D bands, and comparable intra-band couplings on the two quasi-1D bands, all of which are roughly consistent with the qualitative observation in the previous section. We verify that these main features are generic for a broad range of interaction parameters (also see (\ref{eq:V0matrixHelical}) in Sec~\ref{sec:TRSB}), and for spin-orbit coupling smaller than $0.1t$.  Note that in this calculation, $\lambda_\text{SOC}=0.1t$ already represents a rather strong spin-orbit coupling\cite{SOCnote} suggested by recent measurements.\cite{Haverkort:08,Veenstra:14,Fatuzzo:15} The relative inter-band interactions between $\gamma$ and the other two bands depends on $\lambda_\text{SOC}$ and are weaker for smaller spin-orbit coupling.

Following Appendix \ref{app1}, solving the compact gap equation (\ref{eq:oldGapEquation}) which is related to (\ref{eq:V0matrix}), we obtain two attractive solutions, with the leading one given by $\hat{\Delta}\sim(0.33,0.31,0.89)^T$,\cite{othersolution} i.e. comparable gap amplitudes on the three bands similar to what was originally obtained in Ref.~\onlinecite{Scaffidi:14}. In this regard, the $\hat{V}$-matrix encapsulates crucial information about the multi-band character of the superconducting state in the pairing channel under consideration. The noticeable attractive interaction on the $\gamma$-band can be attributed to the proximity to the van Hove singularity on that band. Interestingly, the quasi-1D bands experience repulsive intra-band interactions which disfavor Cooper pairing. This is however compensated by a strong interaction induced by the pronounced incommensurate spin fluctuations between the two bands, which is even stronger than the intra-band interaction on $\gamma$, to make the pairing strengths on the two sets of bands comparable.  In fact, over a wide parameter range ($0<J/U<0.3$) one finds comparable gap magnitudes on all bands.\cite{Scaffidi:14} 


As an important remark, while the one-loop weak coupling calculations likely have captured reasonably well the structure of the interactions and hence the symmetry and structure of the gaps, they could potentially predict inaccurate relative gap amplitudes on the bands. For example, at finite interaction scale, due to the quasi-nesting, the inter-band interactions between the 1D bands can in principle be enhanced once higher order scattering processes, such as at the level of random phase approximation, are included. This would accordingly enhance the pairing on these two bands with respect to that on $\gamma$. Of course, higher order contributions could also enhance the effect of the van Hove singularity, which would have the opposite effect, but this may be mitigated somewhat by the fact that the odd-parity gap function must vanish at the van Hove point. 

In addition, in contrast to the results found here and originally in Scaffidi {\it et al.}~\cite{Scaffidi:14}, two recent numerical functional renormalization group approaches~\cite{Wang:13,Tsuchiizu:15} have reported dominant triplet superconductivity on one of the two sets of bands. However, the two predictions differ in an important manner. Wang {\it et al.}~\cite{Wang:13} argued that the small wavevector spin fluctuations associated with the $\gamma$-band van-Hove singularity dominates the superconducting correlation (see also alternative argument by Huo {\it et al.}\cite{Huo:13}), while in Tsuchiizu {\it et al.}~\cite{Tsuchiizu:15} superconductivity is driven primarily by the large wavevector spin fluctuations associated with the quasi-1D bands. The latter study also found noticeable proximity induced superconductivity on the $\gamma$-band once spin-orbit coupling is included.  While we cannot resolve the on-going debate with our calculations, it may not be ruled out that both the small and large wavevector spin fluctuations enter at low energies with similar strength, promoting comparable pairings on all bands, as is found in this work.

\section{Leggett modes and their detection in Raman spectroscopy}
\label{sec:Raman}
To analyze the collective phase modes, we turn to the Josephson coupling matrix $\hat{\mathcal{V}} = -\hat{V}^{-1}$ using (\ref{eq:V0matrix}), 
\begin{equation}
\hat{\mathcal{V}}=\frac{1}{V_0}\begin{pmatrix}
0.2653 &  0.9302  &  - 0.0734\\
   0.9302&   0.4019 &   -0.0835\\
    -0.0734&    -0.0835&   1.0000
\end{pmatrix} \,. 
\label{eq:Vsro}
\end{equation}
The qualitative features of the inter-band couplings we discussed below (\ref{eq:Vinverse}) are reproduced. The excitation gap of the soft Leggett mode may now be obtained following Appendix \ref{app:Leggett}, 
\begin{equation}
w_{L} \simeq \sqrt{\frac{0.08}{N_0V_0}}\Delta_0  \,,
\label{eq:SROLeggett}
\end{equation}
where we have used the relation $2.8 \Delta_{0\alpha} \simeq 2.8 \Delta_{0\beta} \simeq \Delta_{0\gamma} =  \Delta_0$ obtained above. One may use a rough weak-coupling estimate $N_0V_0 \simeq 0.2$. Thus the energy required to excite this mode ($\sim 0.64\Delta_0$) is lower than the $2\Delta_0$ needed to break a Cooper pair in a fully gapped isotropic superconductor. Nevertheless, due to the strong anisotropy of the superconducting gap structure, low-lying quasi-particles should also exist well below $2\Delta_0$.  Furthermore, with weaker spin-orbit coupling, $\eta_1$ and $\eta_2$  decrease, thus the Leggett mode becomes softer accordingly. Finally, at constant $\Delta_{0\gamma}$, $w_L$ increases (decreases) with increasing (decreasing) $\Delta_{0\alpha,\beta}$. 

We now discuss the experimental consequences for the Leggett modes. These collective phase modes couple indirectly to external electromagnetic fields and hence can be excited by photons in optical probes, such as electronic Raman spectroscopy. The Raman response can be derived via standard linear response theory, and we refer to Refs~\onlinecite{Devereaux:95,Lin:12,Khodas:14,Maiti:15} for details. Essentially, when the frequency difference between the appropriate incident and scattered photons matches the excitation gap of a collective mode, the Raman spectrum exhibits a sharp resonance, as has been observed in the multi-band MgB$_2$ superconductor~\cite{Blumberg:07}.  Moreover, since the Leggett modes correspond only to the relative phase fluctuations between the bands and do not perturb the symmetry of the Cooper pair wavefunction within the individual bands, they couple only to the $A_{1g}$ channel. Thus the Raman spectrum in the $A_{1g}$ channel is a direct measurement of the properties of these phase modes. In realistic situations, the sharp resonances are broadened due to damping effects introduced by impurities and low energy quasi-particles in anisotropic superconductors~\cite{Hirschfeld:92}. 

In addition to the Leggett modes, there exists another form of collective phase fluctuations in chiral superconductors -- the so-called clapping modes~\cite{Wolfle:77}. These modes originate from relative phase oscillations between the two components of the chiral order parameter (thus belonging to angular momentum 2$\hbar$ fluctuations) and are characterized by an excitation gap $\sqrt{2}\Delta_0$ for a two dimensional isotropic chiral $p$-wave superconductor.\cite{Higashitani:00,Kee:00,Kee:03,Sauls:15} The excitation energy may become smaller for anisotropic superconductors.\cite{Sauls:15,Chung:12} Note that by symmetry orthogonal order parameter components from different bands do not couple, so that one can treat the Leggett and clapping modes separately. In principle, the clapping modes also manifest as resonances in optical spectroscopies.\cite{Hirschfeld:92,Higashitani:00,Kee:03,Sauls:15}  However, they do not couple to the Raman $A_{1g}$ channel,\cite{Kee:03} thus distinguishing them from the Leggett modes.

\begin{figure}
\subfigure{ \includegraphics[width=4cm]{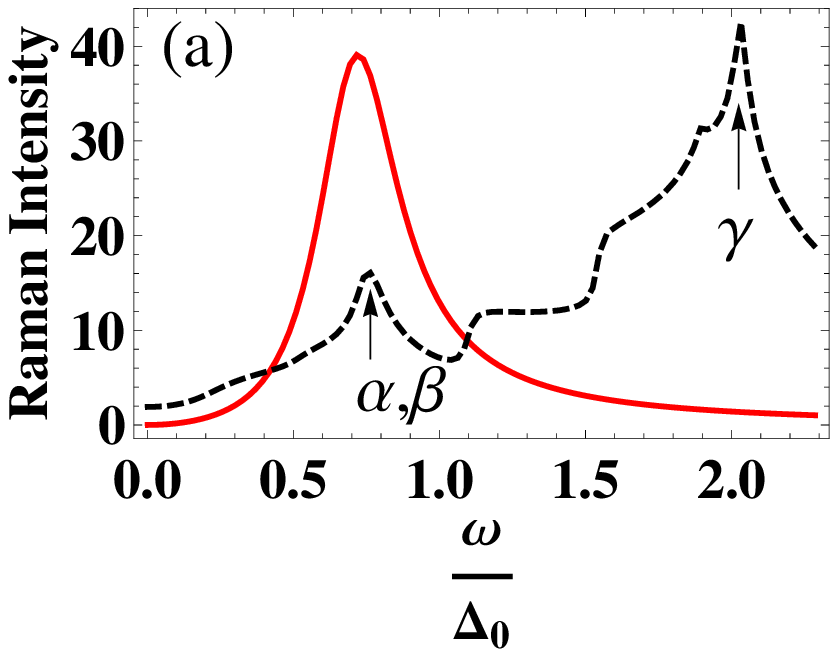} }
\subfigure{ \includegraphics[width=4cm]{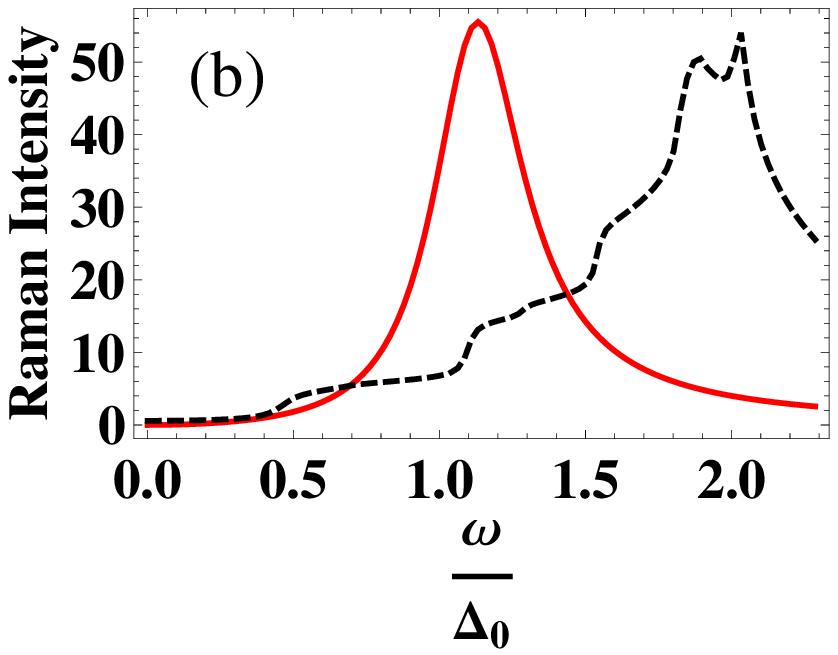} }
\caption{Typical low-frequency electronic Raman response in \SRO~in the $A_{1g}$ channel in the cases of two different gap ratios between the 1D and 2D bands: (a) $|\Delta_{0\gamma}| \simeq 2.8|\Delta_{0\alpha,\beta}|$, (b) $ |\Delta_{0\gamma}| \simeq |\Delta_{0\alpha,\beta}|$. The continuum contribution (black dashed), exhibiting nonvanishing intensity below $\omega=2\Delta_0$, is evaluated using an anisotropic three-band chiral $p$-wave model. The gap anisotropy resembles the one obtained in Ref.~\onlinecite{Scaffidi:14}. A small imaginary part $\tau = 0.004\Delta_0$ is used in the analytic continuation of the Raman susceptibility for regularization. The Leggett mode contribution is shown in red (solid). Here we have simplified the calculation by using a broadening of $\tau_L=0.5\Delta_0$ to model the damping due to the coupling with the low-lying quasi-particle states. This does not alter the Leggett mode peak position~\cite{Hirschfeld:92}.}
\label{fig:Raman}
\end{figure}

Following the derivations in Ref.~\onlinecite{Lin:12}, we plot in Fig~\ref{fig:Raman} typical $A_{1g}$ Raman response for a three-band chiral $p$-wave model of \SRO~using (\ref{eq:Vsro}).  We have considered two different scenarios, one with smaller gap amplitude on the $\alpha$ and $\beta$-bands, another with approximately equal gaps on all bands. In both cases, the Leggett mode manifests as a peak in the spectra. In the cases of sufficiently distinct gaps on the two sets of bands, however, the Leggett mode peak can in principle overlap (as in Fig.\ref{fig:Raman} a), or even switch position with the continuum peak associated with the smaller gap(s), thus potentially complicating the identification of the soft mode. 

In summary, the existence of a low-frequency peak in the $A_{1g}$ channel of the Raman spectrum should be a characteristic feature of the relatively soft Leggett mode in the multi-band \SRO. However, some technical difficulties in Raman spectroscopy, such as laser heating, must be overcome in order to perform measurement at sub-Kelvin temperatures. Furthermore, since the Leggett mode resides at sub-meV frequency range, it may be obscured by the wings of the elastic peak in the Raman spectrum.

\section{Near degenerate order parameters and TRSB}
\label{sec:TRSB}
Here we examine the possibility of TRSB multi-band pairing. This has been predicted for some multi-band (three or more bands) systems when there are two degenerate or near degenerate order parameters, as can occur when the inter-band interactions are frustrated.\cite{Agterberg:99,Voelker:02,Stanev:10,Lin:12,Maiti:13,Biswas:13} In that case, the system may pick a complex linear combination of the two near-degenerate order parameters and, consequently, break time-reversal symmetry. Typically for there to be only a single transition with the TRSB phase condensing at $T_c$, it requires fine tuning and degenerate order parameters. But this phase may exist over a range of parameters (for example as doping is varied), condensing at a second transition below $T_c$.

For the chiral channel of \SRO, the inter-band interactions in (\ref{eq:V0matrix}) are unfrustrated, as they are all attractive, i.e. the three band gaps can choose to have the same sign to simultaneously minimize the inter-band interactions. However, due to the relatively weak interactions between $\gamma$ and the other two bands, the system might still permit two near degenerate solutions, with $\Delta_{0\gamma}$ taking opposite signs with respect to the other bands. 

If two solutions are sufficiently close to degeneracy, whether the two solutions would form a TRSB complex order parameter can then be analyzed using an effective Ginzburg-Landau theory (see Appendix \ref{app:Criterion}). Taking order parameter fields $\Delta_1$ and $\Delta_2$ to denote the respective amplitudes of the leading and subleading solutions, we find that the relative phase between the two is determined by three quartic terms in the free energy: $\beta^\prime |\Delta_1|^2(\Delta^\ast_1\Delta_2 + \Delta_1\Delta_2^\ast)$, $\beta^{\prime\prime}|\Delta_2|^2(\Delta^\ast_1\Delta_2 + \Delta_1\Delta_2^\ast)$, and $\beta (\Delta^\ast_1\Delta_2 + \Delta_1\Delta_2^\ast)^2$ ($\beta>0$), the first two of which favor a non-TRSB real superposition of the two fields, while the last term promotes complex superposition. Since $\Delta_1$ dominates below $T_c$, the $\beta^\prime$ term is most significant. We thus conclude that this type of complex multi-band order parameter is unlikely to develop in our system. 

Another interesting possibility is a TRSB helical state. However, we verify through our microscopic calculations that the inter-band interactions in the helical channel are qualitatively similar to those of the chiral channel (thus our previous discussions of a relatively soft Leggett mode equally applies to the helical channel).  For example, using $J/U=0.08$, $\lambda_\text{SOC} = 0.1t$, as in Ref.~\onlinecite{Scaffidi:14}, we obtain for the helical channel, similar to (\ref{eq:V0matrix}),
\begin{equation}
\hat{V} = V_0
\begin{pmatrix}
    0.6185&   -1.6331 &  -0.0635\\
   -1.6331&    0.5193 &  -0.0677\\
   -0.0635&   -0.0677 &  -1.0000
\end{pmatrix}  \,. 
\label{eq:V0matrixHelical}
\end{equation}
Thus the inter-band interactions are also unfrustrated. The leading attractive order parameter has $\hat{\Delta}\sim (0.50,0.65,0.57)$, i.e. comparable gap amplitudes on the thee bands as in the chiral channel. Combined with the Ginzburg-Landau analysis, we see that the TRSB multi-band pairing is equally unlikely in this channel.

\section{Conclusions}
In this work we have focused on some novel aspects of \SRO~associated with the multi-band nature of the superconductivity in this material. Our qualitative and quantitative analyses yield a consistent description of the multi-band interactions and couplings between the three bands. In particular, in line with an earlier argument,\cite{Agterberg:97} the distinct dimensional characters of the quasi-2D $\gamma$-band and the quasi-1D $\alpha/\beta$-bands in general results in a rather weak coupling between the two sets of bands. Such a peculiar coupling scheme permits a relatively soft Leggett mode, which may be detected in optical probes such as Raman scattering, thereby providing a testing ground for understanding the nature of the multi-band unconventional superconductivity in \SRO. 

In addition, our microscopically evaluated band interactions indicate comparable pairing interactions on the quasi-1D and 2D bands, although from quite different origins, thus clarifying the origin for \SRO~to exhibit comparable gaps on the three bands. We note this is compatible with specific heat measurements\cite{Nashizaki:00} and recent tunneling spectroscopy measurements.\cite{davistunnel}  

We also discussed the possibility of novel TRSB multi-band superconductivity, in both chiral and helical channels. However, \SRO~lacks the frustrated inter-band interactions favorable for the formation of the complex multi-band order parameter. Nevertheless, given the difficulties in reconciling chiral $p$-wave pairing\cite{Maeno:12,sro-review,chiral-review,Matsumoto:99} and the strict experimental upper bounds placed on the edge current,\cite{Kirtley:07,Hicks:10,Curran:14,Liu:16} as well as the indications of Pauli limiting effect in this material,~\cite{Deguchi:02,Yonezawa:13} it might be instructive to investigate the possibility of alternative TRSB superconductivity which does not necessarily involve chiral $p$-wave pairing. 


Finally, although our discussions are focused on \SRO, the analyses are suitable for studying the nature of multi-band superconductivity in other systems. 
\acknowledgements
We would like to thank Lara Benfatto, John Berlinsky, Girsh Blumberg, Andrey Chubukov, Ye-Hua Liu, Yoshi Maeno and Maurice Rice for various helpful discussions. This work is supported in part by NSERC (CK and WH), CIFAR (CK), the Canada Research Chair program (CK), the National Science Foundation under Grant No. NSF PHY11-25915 (CK) and by a grant from the Simons Foundation ($\#$395604 to Catherine Kallin). TS acknowledges the financial support of the Clarendon Fund Scholarship, the Merton College Domus and Prize Scholarships, and the University of Oxford.  MS acknowledges financial support through an ETH Zurich research grant and the Swiss National Science Foundation. TS, CK and WH thank the hospitality of Kavli Institute for Theoretical Physics where part of the work was completed. WH is grateful for the hospitality of the Pauli Center for Theoretical Studies at ETH Zurich where part of the work was completed. 

\appendix
\section{Gap equation and inter-band Josephson coupling}
\label{app1}
In the weak coupling renormalization group calculation presented in Scaffidi {\it et al.},\cite{Scaffidi:14} the effective interaction $V^{\mu\nu}(\bk,\bp)$ that scatters Cooper pairs is obtained by including all of the contributing diagrams up to the one-loop level. Here the wavevectors $\bk$ and $\bp$ are Fermi wavevector on band $\mu$ and $\nu$, respectively. As will be further elaborated below, the ``average" inter-band interaction {\it in a particular pairing channel} is a good measure of the strength of the inter-band Josephson coupling pertaining to that channel. 

One can solve the linearized gap equation to obtain solutions belonging to different pairing channels, 

\begin{equation}
\phi_\mu (\bs k)\Delta_{0\mu} = -C\sum_{\nu=\alpha,\beta,\gamma} \oint_{\nu FS}\frac{d\bs p}{ v_\nu(\bs p)} V^{\mu\nu}(\bs k, \bs p)  \phi_\nu (\bs p)\Delta_{0\nu} \,. 
\label{eq:gap}
\end{equation}
Here $C=\text{ln}\frac{1.13W_D}{T_c}$, $\Delta_{0\mu}$ is the amplitude of the superconducting gap on band-$\mu$, $\phi_\mu(\bk)$ is the normalized form factor characteristic of the symmetry and structure of the gap, and $v_\nu(\bp)$ is the Fermi velocity of band-$\nu$. The most attractive eigen solution of (\ref{eq:gap}) corresponds to the leading superconducting instability with largest $T_c$. 

The Josephson coupling between the bands  in any particular pairing channel may be extracted through the following procedure. First multiply both sides of (\ref{eq:gap}) by $\frac{1}{v_\mu(\bk)}$, perform an integration over $\bk$, and define the following quantities,
\begin{equation}
A_\mu= \oint_{\mu FS} d{\bs k} \frac{\phi^\ast_\mu (\bs k) \phi_\mu (\bs k)}{v_\mu(\bs k)} 
\label{eq:Amu}
\end{equation} 
\begin{equation}
V^{\mu\nu}_0 = \oint_{\mu FS}d\bk \oint_{\nu FS} d\bp \frac{\phi^\ast_\mu (\bs k) V^{\mu\nu}(\bs k, \bs p) \phi_\nu (\bs p)}{v_\mu(\bs k) v_\nu(\bs p)}
\label{eq:V0}
\end{equation}
It is easy to show that $V^{\mu\nu}_0 = (V^{\nu\mu}_0)^\ast$. The gap equation (\ref{eq:gap}) can then be transformed into a simple matrix form, 
\begin{equation}
\begin{pmatrix}
A_\alpha \Delta_{0\alpha} \\
A_\beta \Delta_{0\beta} \\
A_\gamma \Delta_{0\gamma}
\end{pmatrix}
=- C\cdot \hat{V}_0
\begin{pmatrix}
\Delta_{0\alpha} \\
\Delta_{0\beta} \\
\Delta_{0\gamma}
\end{pmatrix}  \,,
\label{eq:oldGapEquation}
\end{equation}
where,
\begin{equation}
\hat{V}_0 = \begin{pmatrix}
V^{\alpha\alpha}_0 &  V^{\alpha\beta}_0  & V^{\alpha\gamma}_0 \\
V^{\beta\alpha}_0 &  V^{\beta\beta}_0  & V^{\beta\gamma}_0 \\ 
V^{\gamma\alpha}_0 &  V^{\gamma\beta}_0  & V^{\gamma\gamma}_0
\end{pmatrix} \, . 
\label{eq:coupling0}
\end{equation}
Eq (\ref{eq:oldGapEquation}) thus constitutes a set of compact gap equations where the form factors of the gap functions are integrated out. Note that all of the eigen solutions of this gap equation belong with the same underlying pairing channel specified by those form factors. In other words, the eigen solutions of (\ref{eq:oldGapEquation}) only give the order parameter configurations (relative amplitudes and signs of the gaps) on the three bands, and the actual gap functions must necessarily contain the characteristic form factors.

We define the effective interactions between the bands as,
\begin{equation}
V^{\mu\nu} = \frac{V_0^{\mu\nu}}{\sqrt{A_\mu A_\nu} } \,.
\label{eq:newV}
\end{equation}
For a one-band system, this returns the effective interaction originally formulated in Ref.~\onlinecite{Scalapino:86}.  Note that if we take a loose approximation $A_\alpha \simeq A_\beta \simeq A_\gamma = A_0$ (which is roughly correct for most of our numerical calculations), the eigen vectors of $\hat{V}_0$ (or $\hat{V}$) constitute the solutions to the gap equation. 

In conjunction with the discussions below (\ref{eq:action}) in the main text, we obtain for the inter-band Josephson coupling, up to an overall constant of the order of the density of states, 
\begin{equation}
\hat{\mathcal{V}} = -\hat{V}_0^{-1}  \,.
\label{eq:coupling}
\end{equation}

As a side remark, the relative signs of the $\Delta_{0\mu}$'s depend on the choice of gauge but the physics remains the same. For example, one can assign an arbitrary sign to the form factor of, say band-$i$, which according to (\ref{eq:V0}) results in a change in the signs of both $V_0^{ij}$ and $V_0^{ik}$ for $i\neq j$ and $i\neq k$. This yields sign changes in the corresponding Josephson couplings $\mathcal{V}^{ij}$ and $\mathcal{V}^{ik}$. However, neither the eigenvalues of (\ref{eq:oldGapEquation}) nor the masses of the collective phase excitations are altered because of the sign change. 

A gauge transformation cannot change the sign of {\it only} one of the three inter-band couplings. Thus one can classify the multi-band superconductivity based on the configuration of the signs of the inter-band Josephson couplings -- a classification beyond the lattice point group symmetries.\cite{Ota:11} For example, $\text{sgn}[\mathcal{V}^{\alpha\beta},\mathcal{V}^{\alpha\gamma},\mathcal{V}^{\beta\gamma} ]=[++-]$ is equivalent to $\text{sgn}[\mathcal{V}^{\alpha\beta},\mathcal{V}^{\alpha\gamma},\mathcal{V}^{\beta\gamma} ]=[+-+]$, as the two can be transformed into one another by changing the sign of the form factor of the $\gamma$-band. 

\section{Leggett modes}
\label{app:Leggett}
Here, we analyze the effective model introduced in Sec~\ref{sec:Model} in detail and highlight the important features of the collective excitations associated with the relative phase fluctuations between the bands. 

We ignore the generically massive order parameter amplitude modes. Making explicit the complex phases of the three gaps, $\Delta_l e^{i\theta_l}$ with $\Delta_l \equiv \Delta_{0l}$ positive real, we can then proceed to derive the dispersion relations for the phase modes, following the standard procedure~\cite{Aitchison:95}. After a gauge transformation, $(\psi_{l\sigma},\psi^\dagger_{l\bar{\sigma}})^T \ra (e^{i\theta_l/2}\psi_{l\sigma},e^{-i\theta_l/2}\psi^\dagger_{l\bar{\sigma}})^T$, the effective action in Eq.~(\ref{eq:action}) becomes,
\begin{equation}
S= \int d\tau d^3r\left[ \sum_{l,j} \Delta_l \hat{\mathcal{V}}_{lj} \Delta_j e^{i(\theta_l-\theta_j)} - \sum_l \text{Tr~ln}(1+\hat{G}_{0l}\Sigma_l)  \right]
\label{eq:action1}
\end{equation}
where the Green's function satisfies,
\begin{equation}
\hat{G}^{-1}_{0l} = -\sigma_0 \partial_\tau + \Delta_l \sigma_1 - \sigma_3 (-\frac{\nabla^2}{2m_l} -\mu_l)
\label{eq:Greenfunction1}
\end{equation}
and the self-energy follows as,
\begin{equation}
\Sigma_l = -\left(\frac{i\nabla\theta_l \cdot \nabla}{2m_l} \right)\sigma_0    + \left[-i\frac{\partial_\tau \theta_l}{2} - \frac{1}{2m_l}\left(\frac{\nabla\theta_l}{2}\right)^2 \right] \sigma_3
\label{eq:selfenergy}
\end{equation}
where $\sigma_\mu$'s are the usual Pauli matrices. 


Consider small amplitude deviations of the phases from the stable state $\theta_l=\theta_{0l}+\phi_l$ , the action in Eq.~({\ref{eq:action1}}) can be expanded with respect to the $\phi_l$'s as~\cite{Lin:12,Ota:11}
\begin{equation} 
S[\phi] = \sum_{n} \int d^3q \hat{\phi}(-w_{n},-q)^T \mathcal{M} \hat{\phi}(w_{n},q)
\label{eq:phaseAction}
\end{equation}
where $\hat{\phi}(w_{n},q) =(\phi_\alpha,\phi_\beta,\phi_\gamma)^T(w_{n},q)$ with $w_{n}= 2n\pi/T $, and the matrix,
\begin{widetext}
\begin{equation}
\mathcal{M} =  \frac{1}{V_0}
\begin{pmatrix}
\mathcal{K}_\alpha-\lambda \epsilon_{\alpha\beta}-\eta\epsilon_{\alpha\gamma} & \lambda\epsilon_{\alpha\beta} & \eta\epsilon_{\alpha\gamma} \\
\lambda\epsilon_{\alpha\beta} & \mathcal{K}_\beta -\lambda \epsilon_{\alpha\beta}-\eta\epsilon_{\beta\gamma} & \eta\epsilon_{\beta\gamma} \\
\eta\epsilon_{\alpha\gamma} & \eta \epsilon_{\beta\gamma} & \mathcal{K}_\gamma-\eta( \epsilon_{\alpha\gamma}+ \epsilon_{\beta\gamma} )
\end{pmatrix} 
\label{eq:M}
\end{equation}
\end{widetext}
with $\mathcal{K}_l = N_l (w_n^2 +\bar{v}^2_{Fl}q^2/2)$, $\epsilon_{lj}= \cos(\theta_{0l}-\theta_{0j})\Delta_{0l}\Delta_{0j}$, where $N_l$ and $\bar{v}_{Fl}$ are respectively the density of states and average Fermi velocity of the $l$-band. 

It is worth noting that the relative phase $\theta_{0l}-\theta_{0j}$, and hence the most stable order parameter configuration, depends on the relative magnitude and signs of the original inter-band interactions. For example, if all inter-band interactions are attractive as in our (\ref{eq:V0matrix}) and (\ref{eq:V0matrixHelical}), the obvious most favorable state has all three order parameters in phase, i.e. $\epsilon_{lj}=1$.  In this case, now consider a rough approximation, $N_\alpha = N_\beta = N_\gamma = N_0$ and $\bar{v}_{F\alpha} =\bar{v}_{F\beta}=\bar{v}_{F\gamma}=\bar{v}_{F0}$, and take the amplitude of the gaps to be the same on all bands. After an analytic continuation, the dispersion relations for the phase modes may be obtained by diagonalizing (\ref{eq:M}),
\begin{eqnarray}
w^2_G &=& \frac{1}{2} \bar{v}^2_{Fl}q^2 \\
w^2_{L1}&=& -3\frac{\Delta_0^2}{N_0 V_0}\eta +  \frac{1}{2}\bar{v}^2_{Fl}q^2 \\
w^2_{L2}&=&  -\frac{\Delta_0^2}{N_0 V_0}(\eta+2\lambda) +\frac{1}{2}\bar{v}^2_{Fl}q^2 
\label{eq:dispersion}
\end{eqnarray}
Here $w_G$ denotes the usual $U(1)$ Goldstone mode, which would be massive had we properly included the vector potential in our formalism; $w_{L1}$ and $w_{L2}$ are the relative phase Leggett modes. Crucially, the excitation gap of the $L1$ mode, as determined by the $\gamma$-$\alpha/\beta$ inter-band Josephson coupling, may be considerably smaller than the superconducting gap. This is the soft Leggett mode we anticipated. Interestingly,  in our case described by, e.g. (\ref{eq:V0matrix}) and (\ref{eq:Vsro}), the $L_2$ mode is overdamped. This mode is related primarily to a relative phase oscillation between the two 1D bands dominated by an inter-band interaction. The readers are referred to Marciani et al.\cite{Marciani:13} for a more extensive discussion of the models containing dominant inter-band interactions. 

Similar analyses carry through when the Josephson couplings take different signs. An interesting scenario arises when the inter-band interactions are frustrated (Sec~\ref{sec:TRSB}), where a much softer Leggett mode is shown to be present~\cite{Lin:12,Ota:11}. From our calculations, this mode is given by $w^2_{L_1}=
\frac{3\Delta_0^2}{2N_0V_0}\frac{|\eta|}{|\lambda|}|\eta|$. To summarize, a relatively soft Leggett mode exists in all scenarios, provided the ground state features comparable pairing gaps on the three bands. 

\section{Two near-degenerate solutions}
\label{app:Criterion}
Here we discuss the scenario where two nearly degenerate attractive order parameter configurations emerge,
\begin{equation}
\hat{\Delta}_{1\bs k} = \Delta_1 \begin{pmatrix}
\eta_{1\alpha} \phi_{\alpha} (\bs k) \\
\eta_{1\beta} \phi_{\beta}( \bs k) \\
\eta_{1\gamma} \phi_{\gamma}( \bs k) 
\end{pmatrix} \,, ~~~
\hat{\Delta}_{2\bs k} = \Delta_2 \begin{pmatrix}
\eta_{2\alpha} \phi_{\alpha}( \bs k) \\
\eta_{2\beta} \phi_{\beta}( \bs k) \\
\eta_{2\gamma} \phi_{\gamma}( \bs k) 
\end{pmatrix}  \, ,
\label{eq:Eigenvectors}
\end{equation}
where $\hat{\eta}_i=(\eta_{i\alpha},\eta_{i\beta},\eta_{i\gamma})^T$ ($i=1,2$) are eigen vectors to (\ref{eq:oldGapEquation}) with eigenvalues $\lambda_1$ and $\lambda_2$ ( assume $|\lambda_1| \gtrsim | \lambda_2|$, $T_{c1} \gtrsim T_{c2}$), and the two fields $\Delta_1$ and $\Delta_2$ describe the amplitude of the superconducting order parameter in the corresponding solutions. In correspondence with the interactions given in (\ref{eq:V0matrix}), for example, $\hat{\eta}_1 =(0.33,0.31,0.89)^T$ and $\hat{\eta}_2= (0.70,0.67,-0.25)^T$. Note we have included in (\ref{eq:Eigenvectors}) the form factors of the gaps on the respective bands for clarity. We are interested in the possibility of TRSB in connection with a complex superposition of the two configurations. This may be best examined within an effective Ginzburg-Landau theory. 

We wish to see if it is favorable for $\Delta_2$ to coexist with the primary $\Delta_1$ below the superconducting transition and form a complex order parameter.  The free energy density reads,
\begin{eqnarray}
f&=& \frac{\alpha_1}{2} |\Delta_1|^2 +  \frac{\alpha_2}{2} |\Delta_2|^2  \nonumber \\
&+& \frac{\beta_1}{4}   |\Delta_1|^4 +  \frac{\beta_2}{4}   |\Delta_2|^4 +  \frac{\beta_{12}}{4}   |\Delta_1|^2|\Delta_2|^2 \nonumber \\
&  +& \frac{\beta^\prime}{4}|\Delta_1|^2(\Delta_1^\ast\Delta_2 + \Delta_1\Delta_2^\ast) +  \frac{\beta^{\prime\prime}}{4}|\Delta_2|^2(\Delta_1^\ast\Delta_2 + \Delta_1\Delta_2^\ast)   \nonumber \\
&+& \frac{\beta}{4} (\Delta_1^\ast\Delta_2 + \Delta_1 \Delta_2^\ast)^2  + ...  \,,
\label{eq:FreeEnergy}
\end{eqnarray}
where ``..." stands for higher order terms, $\alpha_i \sim (T-T_{c,i})/T_{c,i}$, and all of the $\beta$-coefficients can be derived from the microscopic band structure and (\ref{eq:Eigenvectors}). Here $\Delta_1$ and $\Delta_2$ share the same point group symmetry and $U(1)$ symmetry (instead of separate $U(1)$), thus the terms with $\beta^\prime$ and $\beta^{\prime\prime}$ are allowed. 

It can be shown that $\beta_{1,2},\beta_{12}, \beta>0$, while $\beta^\prime$ and $\beta^{\prime\prime}$ can take both signs and are in general non-vanishing. Note in Maiti and Chubukov\cite{Maiti:13}, $\beta^\prime=\beta^{\prime\prime}=0$ due to the particular structure of the eigenbasis resulting from the effective multi-band interactions of their model. In our case for the type of interactions similar to (\ref{eq:V0matrix}) and (\ref{eq:V0matrixHelical}), $\beta^\prime, \beta^{\prime\prime} \neq 0$ and their magnitudes are of the same order as $\beta$. 

Although $\Delta_2$ is not expected to condense right below $T_{c1}$, the coupling between $\Delta_1$ and $\Delta_2$ associated with the $\beta^\prime$ term immediately induces a non-vanishing $\Delta_2$ growing with $\Delta_1$ as $|\Delta_2|\propto |\Delta_1|^3\propto [(T_c-T)/T_c]^{\frac{3}{2}}$. Similarly the third possible order parameter (which is not near-degenerate and not explicitedly written down in (\ref{eq:Eigenvectors})) will mix in below $T_{c1}$, but, in general, with a smaller amplitude. These sub-dominant components grow slower than $\Delta_1$, but this nevertheless suggests that determining the low temperature multi-band gap amplitudes requires going beyond the weak-coupling approximations we used. 

Irrespective of how a non-vanishing $\Delta_2$ may arise below $T_{c1}$, when the two fields coexist, their relative phase is determined by the last three quartic terms in the free energy, of which the $\beta$-term favors a TRSB complex superposition, while the $\beta^\prime$- and $\beta^{\prime\prime}$-terms favor non-TRSB order parameters.  Since $|\Delta_1| \gg |\Delta_2|$, the $\beta^\prime$ term dominates. Thus we conclude that our system is unlikely to sustain a TRSB complex multi-band order parameter.


\begin{thebibliography}{99}
\bibitem{othersolution} The other solutions are $(0.70,0.67,-0.25)^T$ (also with an attractive eigenvalue) and $(-0.77,0.63,0.00)^T$.
\end{thebibliography}
\end{document}